# Trend Analysis of Meteorological Parameters, Tropospheric Refractivity, Equivalent Potential Temperature for a Pseudo-adiabatic Process and Field Strength Variability, Using Mann-Kendall Trend Test and Sen's Estimate.

**[1 & 2, *] Emmanuel P. Agbo, [1] Chris M. Ekpo, [3] Collins O. Edet**

*[1] Department of Physics, Cross River University of Technology, Calabar, Nigeria.*
*[2] Utility Department, Lafarge Africa PLC (A member of LafargeHolcim), Mfamosing Plant, Akamkpa, Cross River State, Nigeria.*
*[3] Theoretical Physics Group, University of Port Harcourt, Port Harcourt, Nigeria.*
*Corresponding author's e-mail: emmanuelpaulagbo@gmail.com*
*Corresponding Author's Postal Code: 540242*
*Corresponding author phone: +2348142525858*
*Corresponding author ORCID: 0000-0002-0215-2492*

**Authors' contributions**

Author 1 designed the study, handled the review, acquired and analyzed the data, author 2 interpreted the data, author 2 and 3 revised the article critically for important intellectual content. All authors read and approved the final manuscript.

## ABSTRACT

Trend analysis of meteorological parameters (temperature, pressure, and relative humidity) as well as calculated refractivity, equivalent potential temperature (EPT) for a pseudo-adiabatic process, and field strength in Calabar, Southern Nigeria has been analyzed using Mann-Kendall trend test and Sen's slope estimator**.** Data of the meteorological parameters were obtained from the Nigerian Meteorological Agency (NiMet) in Calabar for 14 years (2005 - 2018). Results show that the maximum and average temperature, atmospheric pressure, refractivity, EPT and field strength all exhibited a positive Kendall Z value with 2.52, 0.33, 3.83, 0.77, 0.44 and 3.18 respectively which indicated an increasing trend over time, with only maximum temperature, atmospheric pressure and field strength showing a significant increase at 5% (0.05) level of significance, since their calculated p-values (0.012, 0.0001, and 0.001) were less than 0.05. The relative humidity and minimum ambient temperature showed a decrease in trend over time as they both had a negative Kendall Z values (-0.11 and -1.09 respectively), however, together with the average ambient temperature and refractivity, their trend wasn't significant 5% level of significance since their calculated p-values were all more than 0.05. Linear regression, correlation and partial differentiation showed that relative humidity has the most effect on the changes in seasonal refractivity and an indirect relationship with field strength variability. The relationship between EPT and refractivity has been discovered to be very strong and positive. Descriptive statistics has been used to portray the seasonal and annual trend of all parameters.

**KEY WORDS:** Trend analysis; meteorology; radio refractivity; field strength; Mann-Kendall;

## 1. Introduction

Trend analysis can better be used to depict and predict the changing patterns and variability of climatic parameters. This analysis gives a proper knowledge about the changing conditions of the climate and its effects, as well as the analyses of meteorological parameters which include ambient temperature (K), atmospheric pressure (hPa), relative humidity (%), calculated equivalent potential temperature (EPT) in kelvin (K), refractivity (N-unts) and field strength variability in decibel (dB) using Mann-Kendall trend test and Sen's slope estimator including an explanation on the seasonal trend for all above mentioned variations in a year and a relationship between all parameters.

Some various phenomena have a direct effect on refractivity and field strength in the lower atmosphere (troposphere). These variations usually affect the propagation of radio signal. The meteorological parameters stated are the key drivers of these variations which bring about effects like atmospheric refraction, interference, bending, ducting, errors in elevation for the acquisition of radar data (Akpootu and Iliyasu, 2017; Chukwunike and Chinelo 2016).

We will be studying the variations of surface refractivity, EPT for a pseudo-adiabatic process, and the radio signal field strength, all derived from meteorological parameters which will also be studied. Agbo et al. (2020), explained that the radio field strength has a high correlation with surface refractivity, hence, being able to understand its variability is very necessary to aid in predicting the performance of radio wave networks with quintessence especially at very high frequencies (VHFs) (Adedayo 2016; Eichie et al. 2015). The idea of relating EPT for a pseudo-adiabatic process to refractivity is quite novel and a lot of studies have not been made to identify the direct link between both, this studies seeks to find the connection between them

As stated above, meteorological parameters affect the propagation direction of electromagnetic (EM) waves, however, the sources of these EM waves have their propagation frequencies, and these frequencies are directly linked to the extent at which the atmosphere affects their signal propagation. Hence, the state of the atmosphere (affected my meteorological parameters) by which these waves propagates is not the sole contributor to the effect but since it is stable, and cannot be mechanically controlled like the frequency and power, studying its variation necessary for better understanding (Adediji, et al. 2014).

Knowing this, we can discern the lower part of the atmosphere (the troposphere), at least, to be like a glass prism, having an index of refraction, n, a little denser than free space (vacuum), hence the refraction (Agbo et al. 2020).

The non-parametric Mann-Kendall (M-K) test and Sen's slope Estimator (SSE) has been used by researchers (Garilov et al. 2018; Kisi and Ay., 2014; Alhaji et al. 2018; Roboaa and Al-Barazanji, 2015; Salmi, et al. 2002; Chattopadhyay et al. 2012; Atta-ur-Rahman and Dawood 2016; Mondal et al. 2012) for applications in various research areas including air temperature fluctuations, water quality parameters, rainfall, atmospheric pollutants, and tropospheric ozone, etc., in different locations around the world. However, we'll use this method together with Sen's slope estimator to analyze the annual trends of the eight distinct parameters.

Most literature (Agbo, et al. 2020; Adedayo. 2016; Agbo et al. 2013; Ajayi and Kolawole, 1984; Tanko et al. 2019; Adediji et al. 2014; Adediji et al. 2011; Adeyemi and Emmanuel 2011; Akinwumi 2018; Kim et al. 2014) have used these meteorological parameters to calculate radio refractivity using the method recommended by the International Telecommunication Union (ITU), but this study has analyzed annual and seasonal variations for each meteorological parameter, calculated EPT, radio refractivity and the variability of the field strength and the link between all of them, by adopting the M-K test, SSE, linear regression, etc.

As a method of analyzing all meteorological parameters, we take note of the EPT for a pseudo-adiabatic process. This takes note of the absolute temperature, the atmospheric pressure, the mixing ratio and the saturated mixing ratio of the …. (which takes note of the relative humidity), and the absolute temperature at the lifting condensation level.

Accurate approximations of the EPT for a pseudo-adiabatic process have been adopted from (Bolton 1960). The link between this and refractivity has been analyzed to show the relationship between the two and how they both relate to the meteorological parameters. The M-K Test will be adopted to analyze the nature trend to show the significance of the increase or decrease as the case may be. The link between the refractivity and equivalent potential temperature is novel, and has not been studied in detail in climatology. Understanding the relationship between them, and with the inclusion of field strength, the influence on radio wave propagation can be understood with facts.

Results from the Mann-Kendall (M-K) test can determine the nature of the trend (either increasing or decreasing) by calculating the Kendall Z and S statistic values; we can discern if it's an increasing trend if the values are positive and a decreasing trend if the values are negative.

The significance of the increasing or decreasing trend as the case may be, can be determined by comparing the significance level (alpha) to the probability value (p-value) of the data observations. If the p-value is less than the significance level, this shows that the null hypothesis is rejected and there is a trend in a series. Similarly, if the p-value is greater than the significance level, this shows that the null hypothesis $H_o$ is accepted and there is no trend in a series. (Alhaji et al. 2018)

Alhaji et al (2018) utilized the Mann-Kendall trend test analyze the variation of temperature in Gombe state of Nigeria. Their results showed that all the studied temperature parameters (maximum, average, and minimum temperature) all had a positive Kendall Z-value (4.38, 4.43 and 1.59 respectively) showing a positive trend, but only the maximum and average temperature trend has a trend that was significant after their p-values was calculated to be 0.0001, which was less that the significance level of 5% (0.05). The minimum temperature trend, although positive, didn't show a significantly increasing trend after having a p-value of 0.107 which was less than the 5% level of significance (0.05).

This approach has been used for various studies together with SSE to estimate the regression of the Kendall's Tau. Introduced by Sen (1968), the SSE has proven useful in research by estimating the best slope for the regression. This current study will be using this technique to analyze the trend of the mentioned parameters for fourteen (14) years in the Calabar, Southern Nigeria because of its applicability for time series distributions.

We start by reviewing the theory of refractivity as it relates to field strength and all obtained meteorological parameters, equivalent potential temperature (EPT) as it relates to the temperature at the lifting condensation level, mixing ratio, etc., and finally the analysis technique; the M-K test and relating it to the Sen's Slope Estimator. The study area is described, as well as the method of data collection and the data processing and analysis method. Annual variations of meteorological parameters (maximum ambient temperature, minimum ambient temperature, average ambient temperature, relative humidity, and atmospheric pressure), refractivity, field strength variability, and EPT for a pseudo-adiabatic process are analyzed using the M-K test. Seasonal variations for all are also plotted. The link between refractivity and all meteorological parameters, as well as the equivalent potential temperature for a pseudo-adiabatic process is discerned with the trend test, correlation, linear regression and theoretical proof.

## 2. Theoretical Background

### 2.1 Theory of Refractivity

One key feature of EM waves is that it obeys the inverse square law which states that an EM wave with a given power density is inversely proportional to the distance-squared from the source. This means that if the distance from the source (for e.g., transmitter) is tripled, the value of the radiated wave in the new location is reduced to one-ninth of its original value (Tanko et al. 2018). These waves can undergo reflection, diffraction, etc. and this will affect its transmission. (Javeed et al. 2018; Oyedum 2008).

Our basic mathematical expression for refractive index can be related to the velocity using the relationship (Javeed *et al*. 2018; Oyedum 2008);

$$n = \frac{V_f}{V_m} \tag{1}$$

where;

$V_f$ is the signal velocity in free space (ms$^{-1}$)

$V_m$ is the velocity with respect to the specified medium (ms$^{-1}$)

The atmosphere is not a perfect vacuum; it has some elements / parameters that impedes and changes the direction of wave propagation, but the refractive index here is almost unity (Edet *et al*. 2017b, 2017c), this brings about some complications when trying to analyze. At the lower atmosphere (troposphere) these values are approximately between 1.00025 to 1.0004, based on the region (Chinelo and Chukwunike 2016), these values reduce to absolute unity in the upper atmosphere.

The term radio refractivity $N$ (scaled up in parts per million 'ppm') is introduced and related to the refractive index n by the relationship below. (Akpootu and Iliyasu 2017; Ojo et al. 2015);

$$n = 1 + N \times 10^{-6} \tag{2}$$

Simplified to;

$$N = (n-1) \times 10^6 \ (N - Units) \tag{3}$$

Refractivity $N$ is dimensionless, although it is expressed in N-units.

Expressing $N$ in terms of our local meteorological parameters (relative humidity, atmospheric pressure and ambient temperature) has been made possible by recommendations from the International Telecommunication Union (ITU). The Union had approved the relationship to expressed as (Akpootu and Iliyasu 2017; Kaissasou et al. 2015; Adediji et al. 2011; Faladun and Okeke, 2013; Alam et al. 2016, Alam et al. 2017; Emerete et al. 2015);

$$N = 77.6 \frac{P}{T} + 3.73 \times 10^5 \frac{e}{T^2} \left( N - units \right) \tag{4}$$

Reduced to;

$$N = \frac{77.6}{T} \left( P + 4810 \frac{e}{T} \right) \left( N - units \right) \tag{5}$$

where

$P$ is the Atmospheric Pressure (hPa)

$e$ is the Atmospheric Vapour Pressure (hPa)

$T$ is the Absolute Temperature (K)

From equation (4) we can observe that refractivity can be expressed as a summation of two components, the dry and wet term. Results regularly show that the contribution of the dry term to the total value of refractivity is always higher that the wet term with the percentage depending on the region. (Edet *et al*. 2017a, 2017b, 2017c; Agbo et al. 2020).

From equation (4) we can observe the dry term to be (Dairo and Kolawole 2017, Ayatunji et al, 2011; Alimgeer et al., 2018):

$$N_{dry} = 77.6 \frac{P}{T} \left( N - units \right) \tag{6}$$

The dry term above is stable and can be gotten from measurements of ambient temperature and atmospheric pressure. However, the polar nature of the water molecules in the troposphere is the driving force behind the wet term variation (Edet *et al*. 2017a, 2017b, 2017c; Agbo et al. 2020; Gao et al. 2008; Igwe et al. 2012);

$$N_{wet} = 3.73 \times 10^5 \frac{e^2}{T^2} \left( N - units \right) \tag{7}$$

Relative humidity $H$, is a huge contributor to the atmospheric vapour pressure, $e$. They are related by (Chukwunike and Chinelo 2016; Segun et al. 2018; Segun et al. 2013);

$$e = \frac{e_s H}{100} \left( hPa \right) \tag{8}$$

where

$e_s$ is the Saturated or maximum vapour pressure (hPa)

$e_s$ is calculated using Clausius-clapeyron relation and is given by (Edet *et al*. 2017a, 2017b, 2017c);

$$e_s = 6.11 \exp \left( \frac{17.26 \left( T - 273.16 \right)}{T - 35.87} \right) \left( hPa \right) \tag{9}$$

where

T is the Absolute temperature (K)

It is also important to note that the radio refractivity $N$, can be written from equation (5) as;

$$N = N_{dry} + N_{wet} \tag{10}$$

Both the dry and wet terms decrease with height above the troposphere to the upper atmosphere. Although they both decay with increasing height, they do so at different rates which will lead to a bi-exponential model (Tanko et al. 2018). The dry term's contribution to the overall value of $N$ always has a higher percentage as explained, however, its contribution to the overall variation of refractivity is always almost constant. The wet term however contributes a lower percentage of the overall value of $N$, but is responsible for most of the variation (variability) of $N$.

We get a mathematical equation to discern this variability. To get the monthly range of refractivity, we get the maximum $N_{s(MAX)}$ and minimum $N_{s(MIN)}$ values of refractivity by using the values from equation (4) to have; (Tanko et al. 2018)

$$Monthly\ Range = N_{s(MAX)} - N_{s(MIN)} \qquad (11)$$

The field strength variability (FSV) is obtained from;

$$FSV = \left(N_{S(MAX)} - N_{S(MIN)}\right) \times 0.2dB \qquad (12)$$

A factor of *0.2dB* is multiplied to the monthly range to give the *FSV* for every change in surface refractivity, this is true for a frequency range of 30 to 300MHz (Adediji 2014; Tanko et al. 2019).

## 2.2 Equivalent Potential Temperature for a Pseudo-Adiabatic Process

To fully understand the equivalent potential temperature (EPT) or potential equivalent temperature $\theta_E$, we need to come into the understanding of factors like; the temperature at the lifting condensation level $T_L$, mixing ratio $r$, potential temperature, equivalent temperature, saturated mixing ration $r_s$ etc.

In the atmosphere, there is always a movement of air with respect to altitudes, and the variation of atmospheric pressure affects the changes in the state of these air parcels, contributing ultimately to radio wave propagation. Condensation is a very important factor relating these changes in pressure/altitude. For an air parcel that remains unsaturated (lacks the maximum amount of water vapour) at a temperature, there will be no condensation at that temperature. The saturated air parcel on the other hand, can condense making it irreversible Bryan (2008).

Considering a pseudo-adiabatic (saturated) process will help us discern the changes that takes place in air parcels. The water released from the condensation for saturated parcels, gets released with heat; this heat is released via the latent heat of condensation/vaporization. We know that in an adiabatic process, heat is not lost or gained in the process, (i.e., temperature does not change) but work is done. Therefore, this heat that is lost via condensation gets added back to the parcel. This means that an unsaturated air parcel cools at a faster rate than a saturated air parcel that cools down slowly.

We take note of the of the absolute temperature at the lifting condensation level $T_L$; this is the temperature in which a parcel of air would rack up if lifted adiabatically to its condensation level. The lifting condensation level is the height at which an air parcel becomes saturated when it is lifted dry adiabatically.

For accurate values of $T_L$ with respect to our location of study, Bolton (1960) gives the relation to be;

$$T_L = \frac{2840}{3.5InT_k - Ine - 4.805} + 55 \qquad (13)$$

*e* is the mixing ration in g⁻¹kg⁻¹

$T_k$ is the absolute temperature in kelvin

The mixing ratio can be gotten from the relation;

$$r = \varepsilon \frac{e}{P-e} \tag{14}$$

where $\varepsilon = \dfrac{R_d}{R_v} = 621.97$

$$r = 621.97 \frac{e}{P-e} \tag{15}$$

$e$ is the vapour pressure calculated from equation (8), and $P$ is the atmospheric pressure observed at the station.

The equivalent potential temperature (EPT) $\theta_E$ is related to the temperature at the lifting condensation level $T_L$ the potential temperature $\theta$, the mixing ratio $r$, the absolute temperature (in kelvin) $T_k$, and the atmospheric pressure P by;

$$\theta_E = T_K \left( \frac{1000}{P} \right)^{0.2854\left(1-0.28\times10^{-3}\,r\right)} \times \exp\left[ \left( \frac{3.376}{T_L} - 0.00254 \right) \times r \left( 1 + 0.81\times10^{-3}\,r \right) \right] \tag{16}$$

Where the factor; $T_K \left( \dfrac{1000}{P} \right)^{0.2854\left(1-0.28\times10^{-3}\,r\right)} = \theta$ is the potential temperature

This is that temperature at which unsaturated air parcel would have if it is adiabatically taken dry to 1000hpa.

We take into account the variations of $C_{pd}$ (specific heat at constant pressure) with temperature and pressure.

We can define $\theta_E$ to be the final temperature in which a parcel of air will attain when it is lifted dry adiabatically (without loss or gain of heat) to its lifting condensation level, $T_L$, therefore making all the moisture condense pseudo-adiabatically (with respect to water saturation), then dropping out the condensed water as it is formed, then finally being brought down dry adiabatically to 1000hPa. The main difference between the equivalent temperature and the equivalent potential temperature (EPT) is that in the case of the equivalent temperature, the air parcel is not brought down adiabatically to 1000hPa. Therefore, the EPT is the temperature an air parcel attains when the equivalent temperature at the initial level is brought back down to 1000hPa.

### 2.3  Mann-Kendall Trend Test

Mann-Kendall trend test is usually used to analyse data that involves time series; it is mostly used for environmental and hydrological data. It is a non-parametric test that does not require conformity to a particular distribution of the data analysed (Alhaji *et al.* 2018). In this test, the data values are not always compared, instead what is being compared is the relative magnitudes of the sample data.

The Mann-Kendall trend test has some advantages when used. As stated, the test is non-parametric, and it doesn't require a normally distributed data, and the sensitivity of the test due to an inhomogeneous series resulting to abrupt breaks is very low (Alhaji et al. 2018).

The application of this test is related to cases where the given a range of data $x_i$ agrees with the relation (Salmi et al. 2002);

$$x_i = f(t_i) + \varepsilon_i \tag{17}$$

$f(t_i)$ here is a function of time which is continuously increasing or decreasing monotonically, $\varepsilon_i$ are the range residuals with zero mean, The variance of the distribution is assumed to be a constant of time.

We have 2 hypothesis which will be tested, the null hypothesis $H_o$, and the alternative hypothesis, $H_1$. $H_o$ signifies no trend, this explains the random ordering of the given a range of data $x_i$ in time (t). $H_1$ signifies that there is a monotonically increasing or decreasing trend. We use the S statistics for a time series with less than 10 data points, and the Z statistics (normal approximation) is used for times series with 10 or more data points.

### 2.3.1 Number of data values less than 10

The Mann-Kendall test statistic S is calculated using the formula (Kisi and Ay. 2014; Alhaji et al. 2018; Salmi, et al. 2002; Chattopadhyay et al. 2012; Atta-ur-Rahman and Dawood 2016)

$$S = \sum_{k=1}^{n-1} \sum_{j=k+1}^{n} \text{sgn}(x_j - x_k) \tag{18}$$

where;

$$\text{sgn}(x_j - x_k) = \begin{cases} +1; & \text{if } (x_j - x_k) > 0 \\ 0; & \text{if } (x_j - x_k) = 0 \\ -1; & \text{if } (x_j - x_k) < 0 \end{cases} \tag{19}$$

$n$ is the number of data values in the studied series. The test can handle missing values and this creates a unique situation where number of years is greater than the number of data values n. However, the test can handle this anomaly.

An upward trend is indicated by a positive $S$ value, and a downward trend is indicated by a negative $S$ value. However higher positive values of $S$ suggest an increasing trend, and lower negative values suggests a decreasing trend. To properly and statistically quantify the significance of the increasing or decreasing trend, we will explain in the next sub-heading the method for computing the probability associated with $S$.

In the M-K test uses four (4) different significance levels α: 10% (0.01), 5% (0.05, 1% (0.01), 0.1% (0.001). Each of these levels have their specified minimum values of n represented in Table 1.

| Significance level ($\propto$) | Required n |
|---|---|
| 0.1 | $\geq 4$ |
| 0.05 | $\geq 5$ |
| 0.01 | $\geq 6$ |
| 0.001 | $\geq 7$ |

**Table 1.** Significance level ($\propto$) required for given numbers of data

The classification of this probability/significance level is important because results can be confused to be entirely true. To discern this, we need to understand that the significance level of say 0.05, means that there is a 5% probability that a mistake will be made while rejecting the null hypothesis $H_o$. Similarly, a significance level of 0.01 means that there is a 1% probability that a mistake will be made while rejecting $H_o$ (Salmi et al. 2002).

### 2.3.2 Number of data values more than 10

We use the normal approximation (Z statistic) if the number of data values $n$ is from 10 and above. But we should note that when $n$ is close to 10, the validity of this approximation might be reduced because of tied or equal values. To find the variance of $S$ '$VAR(S)$', we compute the equation below. This equation takes the ties that may be present the tied or equal values that are in the time series. (Kisi and Ay. 2014; Alhaji et al. 2018; Salmi, et al. 2002; Chattopadhyay et al. 2012; Atta-ur-Rahman and Dawood. 2016)

$$VAR(S) = \frac{1}{18}\left[ n(n-1)(2n+5) - \sum_{p=1}^{g} t_p (t_p - 1)(2t_p + 5) \right] \tag{20}$$

From the equation, the number of data values is represented by $n$, the number of equal of tied groups is represented by $g$, and the number of data values in the $p^{th}$ group is represented by $t_p$.

The test statistic $Z$ is computed using the values of $VAR(S)$ and $S$; (Kisi and Ay. 2014; Alhaji et al. 2018; Salmi et al. 2002; Chattopadhyay et al. 2012; Atta-ur-Rahman and Dawood., 2016)

$$Z = \begin{cases} \dfrac{S-1}{\sqrt{VAR(S)}}; & S > 0 \\ 0; & S = 0 \\ \dfrac{S+1}{\sqrt{VAR(S)}}; & S > 0 \end{cases} \tag{21}$$

The probability density function for a normal distribution is given by; (Kisi and Ay. 2014; Alhaji et al. 2018; Salmi, et al. 2002; Chattopadhyay et al. 2012; Atta-ur-Rahman and Dawood 2016)

$$f(z) = \frac{1}{\sqrt{2\pi}} e^{-\frac{z^2}{2}}$$

(22)

This above function is with mean of 0 (zero), and a standard deviation of 1 for a normal distribution. We have a decreasing trend if the value of the test statistic $Z$ is negative and the computed p-value (probability from the data values) is higher than the level of significance. Similarly, we have an increasing trend if the $Z$-value is positive and the computed p-value (probability from the data values) is higher than the level of significance. In the special case where the positive or negative $Z$-values are not signifying trends, we discern this from the observation that the computed p-value (probability from the data values) is less than the level of significance.

### 2.4 Sen's Slope Estimator

To estimate the slope of a linear existing trend, the nonparametric Sen method is used. We represent the linear equation by (Alhaji et al. 2018; Sen 1968)

$$f(t) = Qt + B$$

(23)

Here, $f(t)$ is described in equation (17) as a function of time which is continuous increasing or decreasing monotonically and $Q$ is the slope and $B$ is a constant. We get the slope estimate in equation (23) by calculating all the slopes of all data value pairs (Sen 1968);

$$Q_i = \frac{x_j - x_k}{j - k}$$

(24)

Here, $j > k$.

If we have n values $x_j$ in the time series, we get as many as $N = n(n-1)/2$ slope estimates $Q_i$. The Sen's estimator of slope is the median of these $N$ values of $Q_i$. The $N$ values of $Q_i$ are ranked from the smallest to the largest and the Sen's estimator is; (Kisi and Ay. 2014; Alhaji et al 2018; Salmi et al., 2002; Chattopadhyay et al. 2012; Atta-ur-Rahman and Dawood. 2016; Sen 1968)

$$Q = \begin{cases} Q_{[(N+1)/2]}; & if\ N\ is\ odd \\ \frac{1}{2} Q_{[N/2]} + Q_{[(N+2)/2]}; & if\ N\ is\ even \end{cases}$$

(25)

A 100(1- α) % two-sided confidence interval about the slope estimate is obtained by the nonparametric technique based on the normal distribution. The method is valid for n as small as 10 unless there are many groups.

### 3. Methodology

### 3.1 Study Area

Located on latitude 4º57'06"N and longitude 8º19'19"E, Calabar (Fig. 1), which is the capital of Cross River State is located in Southern Nigeria. She has an approximate land area of 157 square miles (406 sq km).

The region is being divided into 2 local government areas called Calabar Municipality and Calabar South. She's a coastal area and has the tropical monsoon climate. The region is elevated about 42m above the sea level. Her closeness to the Atlantic Ocean (coastal area) drives conventional waves to the area, making the area very humid. These waves bring about high humidity which is directly related to rainfall, as studied by (Agbo et al. 2020).

Most of the months in the region experience precipitation apart from the very dry months. The wet season in the region spans over a long period of about 7-10 months and due to the lengthy dry season, there is a stability in the weather conditions throughout the year.

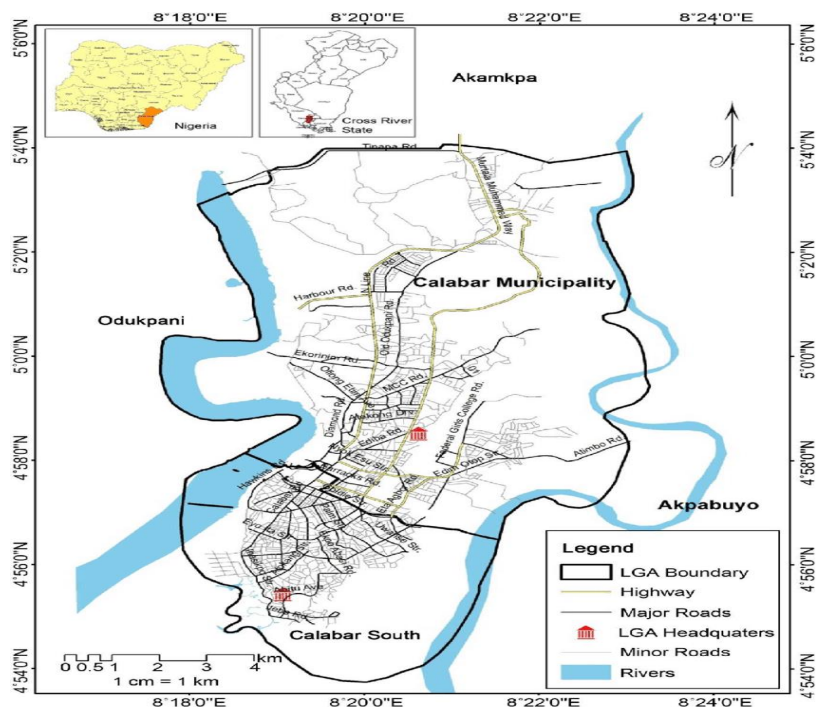

**Fig 1** Map of the Study Area (Agbo et al. 2020)

### 3.2  Data Acquisition

Obtained from the archive of the Nigerian Meteorological Agency (NiMet) in Calabar, the diurnal minimum and maximum ambient temperature (K), the relative humidity (%) and the atmospheric pressure (hPa) for fourteen (14) years were used for this study. NiMet Calabar is located at the Margaret Ekpo International Airport, Calabar (Fig. 2). The hourly data is originally measured and then the average is found to get the diurnal data, these covers all days and seasons of the year. All these were computed at the station.

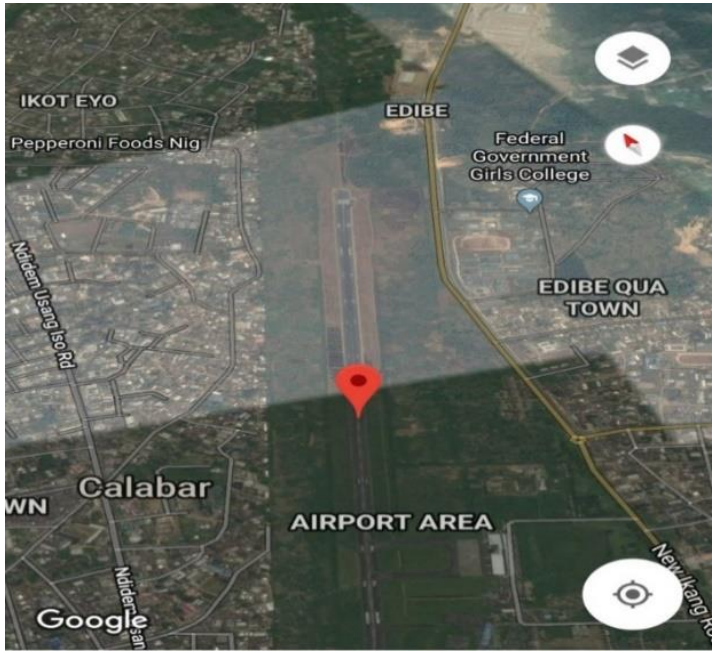

**Fig. 2** Calabar metropolis showing the location of NIMET at Margaret Ekpo International Airport (Agbo et al. 2020).

### 3.3 Data Processing Method

The diurnal minimum and maximum temperature, relative humidity and atmospheric pressure were all obtained from NiMet. The average ambient temperature is calculated from the averaging minimum and maximum temperature. These diurnal values were averaged to get the monthly values from these parameters. Radio refractivity has been calculated using the indirect method recommended by the International Telecommunication union (ITU). To directly measure radio refractivity, refractometers is the most reliable method, but very costly.

Refractivity is calculated by substituting the obtained values of the meteorological parameters into equation (4). Using the calculated monthly refractivity from equation (4), the field strength variability (FSV) has been computed by adopting equation (11) and (12), the EPT has been computed using equation (16). Linear regression equations are calculated and presented to show the degree of influence on each meteorological parameter, and EPT on radio refractivity as well as the coefficient of correlation for each. The M-K test and Sen's estimate has been used to study the time series variation. The test is conducted using an Excel add-in software called "XLSTAT" used for statistical analysis, some results like the Kendall Z-value were computed without the software by using equation (21).

The average ambient temperature, relative humidity, atmospheric pressure data were used to analyze the changes in trend happening in Calabar over the 14 years under study. The measured values of annual refractivity, field strength variability, and equivalent potential temperature were analyzed using Mann-Kendall test too. Variations of the seasonal trends were also analyzed using each parameter for months in a year.

Results have been interpreted with graphs, tables, modelled equations and ultimately descriptions to infer changes in trends. The reader can draw more results relating to climatic changes and using modelled equations. These processes can be applied to other regions.

XLSTAT 2020 has been used for the M-K statistical test to detect the trends in the data points. The null hypothesis ($H_0$), which assumes that there is no trend in the series is tested against the alternative hypothesis ($H_1$) which assumes that there is a trend in the series. The results which informed these hypotheses were the as a result of comparing the probability values of the data to the significance level.

The LINEST function in Excel has been used to calculated the required results of the linear regression between refractivity and the meteorological parameters, these results were later interpreted by modelled liner equations. The coefficient of correlation is then obtained for all parameters to determine their effects on refractivity as well as the equation relating each parameter with refractivity.

## 4 Results and Discussion

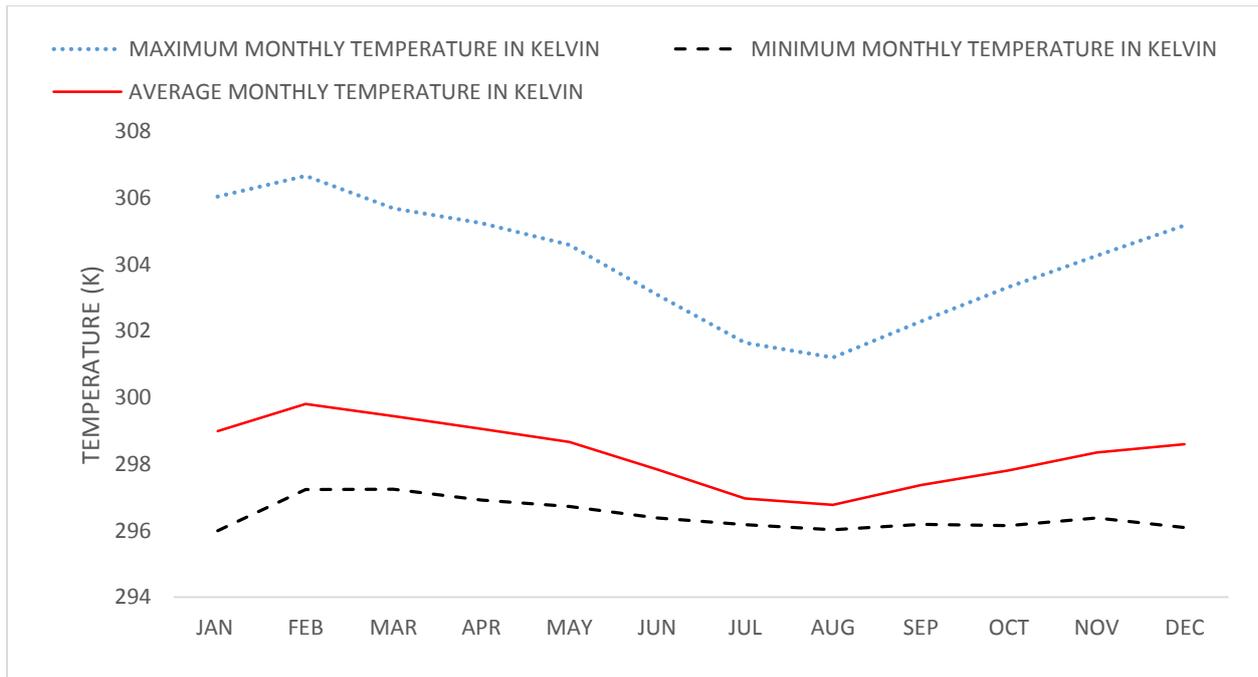

**Fig. 3** Mean seasonal minimum, maximum, and average ambient temperature variation over the period of 14 years

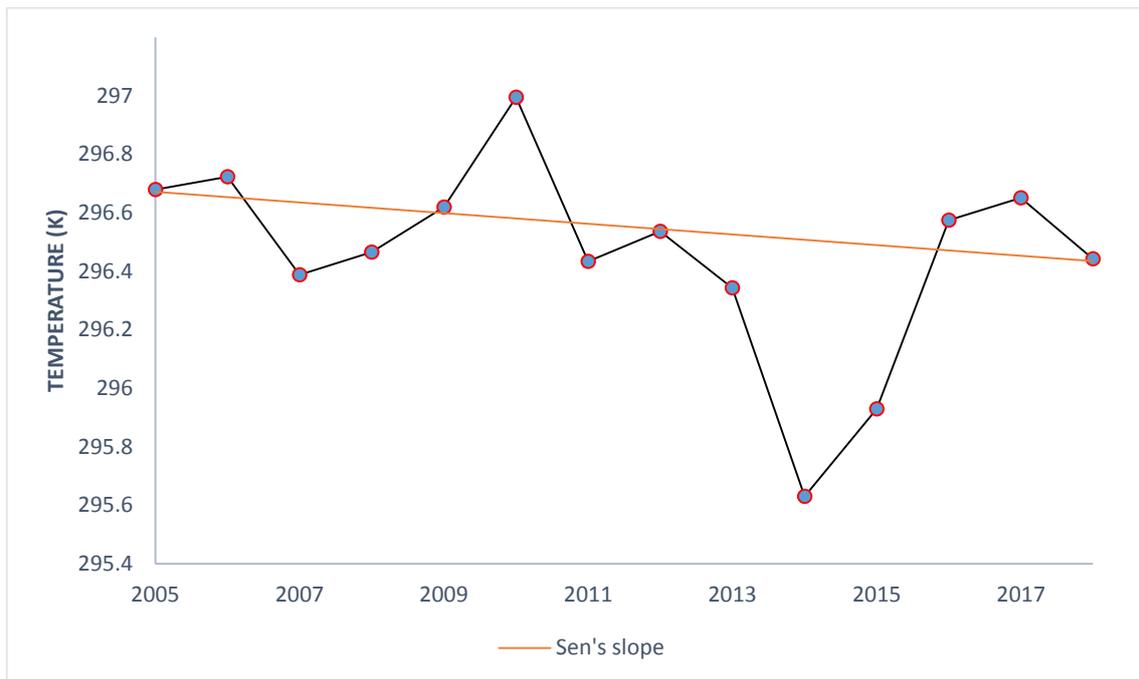

**Fig. 4a** Sen's slope of minimum ambient temperature for the period of 2005-2018

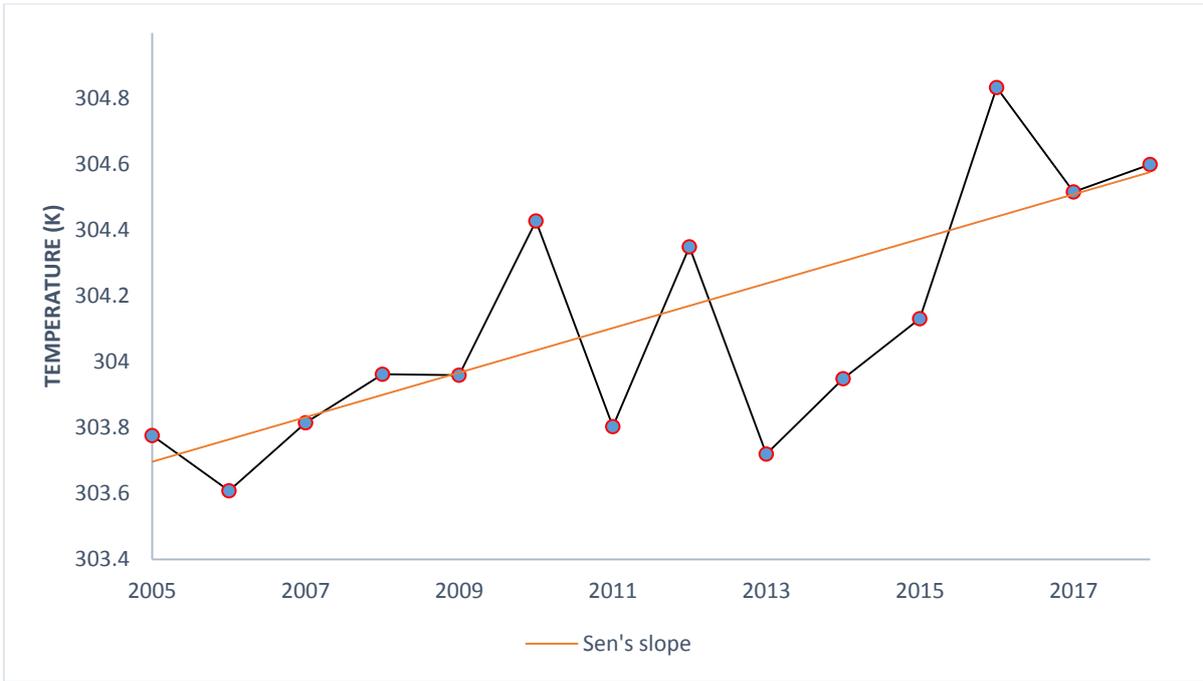

**Fig. 4b** Sen's slope of maximum ambient temperature for the period of 2005-2018

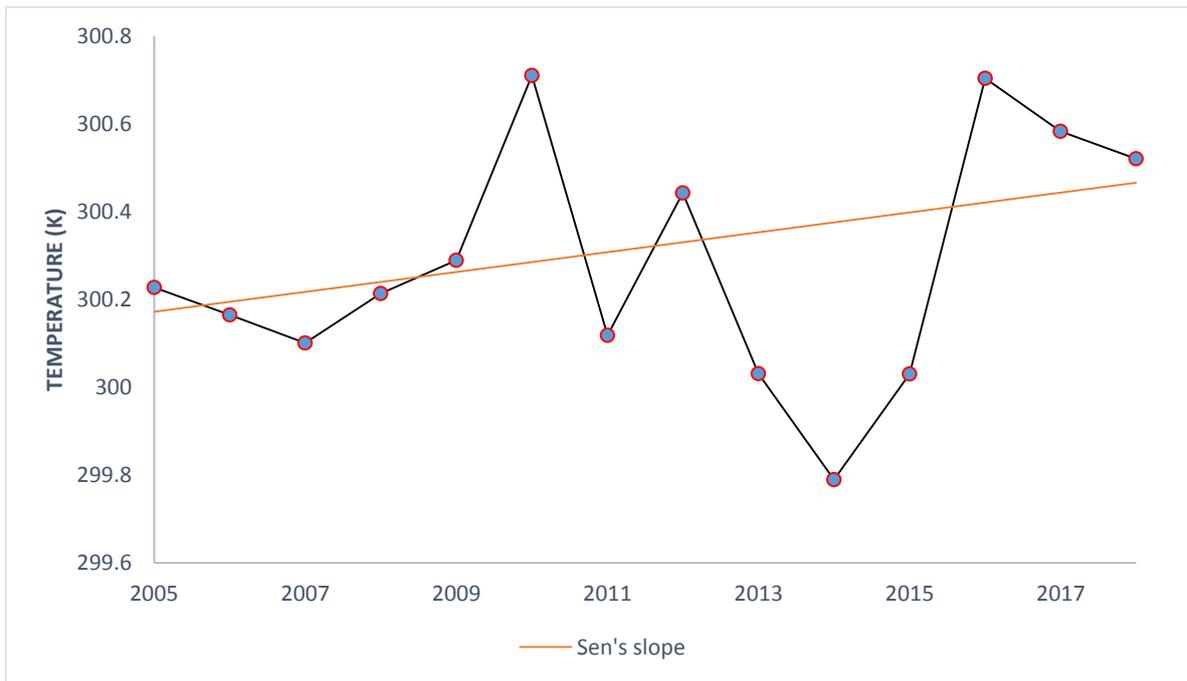

**Fig. 4c** Sen's slope of average ambient temperature for the period of 2005-2018

|  | Minimum (month) | Maximum (month) | Range | Std. Deviation | Mean |
|---|---|---|---|---|---|
| Maximum monthly temperature (K) | 301.21 (August) | 306.66 (February) | 5.46 | 1.78 | 304.10 |
| Minimum monthly temperature (K) | 296.02 (August) | 297.25 (March) | 1.25 | 0.46 | 296.46 |
| Average monthly temperature (K) | 296.78 (August) | 299.81 (February) | 3.03 | 0.97 | 298.31 |
| Relative humidity (%) | 70.63 (January) | 91.81 (August) | 21.18 | 6.98 | 85.71 |
| Atm. Pressure (hPa) | 1004.52 (March) | 1007.82 (June) | 3.30 | 1.24 | 1005.97 |
| Refractivity (N-units) | 368.08 (January) | 396.03 (March) | 27.95 | 8.55 | 387.11 |
| EPT (K) | 350.13 (January) | 364.57 (March) | 14.44 | 4.68 | 357.37 |
| Field Strength (dB) | 5.96 (August) | 10.86 (February) | 4.89 | 1.65 | 8.82 |

**Table 2a** Descriptive statistics for all obtained meteorological parameters, calculated equivalent potential temperature (EPT), refractivity and field strength for months of the year showing the months with the minimum and maximum values of each parameter, mean, range and standard deviation.

| Variable | Observations | Minimum (Year) | Maximum (Year) | Range | Mean | Std. deviation |
|---|---|---|---|---|---|---|
| Field Strength (dB) | 14 | 7.85 (2006) | 9.73 (2016) | 1.88 | 8.82 | 0.57 |
| Refractivity (N-units) | 14 | 383.38 (2014) | 391.57 (2012) | 8.19 | 387.11 | 2.44 |
| EPT (K) | 14 | 354.59 (2014) | 359.92 (2012) | 5.3 | 357.36 | 1.5 |
| Atmospheric pressure (hPa) | 14 | 1005.15 (2005) | 1007.09 (2017) | 1.94 | 1005.97 | 0.66 |
| Average annual temperature (K) | 14 | 299.79 (2014) | 300.71 (2010) | 0.92 | 300.28 | 0.28 |
| Minimum annual temperature (K) | 14 | 295.63 (2014) | 296.99 (2010) | 1.36 | 296.46 | 0.34 |
| Maximum annual temperature (K) | 14 | 303.61 (2006) | 304.83 (2016) | 1.23 | 304.10 | 0.38 |
| Relative humidity (%) | 14 | 83.26 (2010) | 87.97 (2012) | 4.71 | 85.71 | 1.43 |

**Table 2b** Annual descriptive statistics for all obtained meteorological parameters, calculated refractivity, equivalent potential temperature (EPT), and field strength showing the years with the minimum and maximum values of each parameter, mean, range and standard deviation.

## 4.1 Ambient Temperature

Mann-Kendall (M-K) test and Sen's Slope Estimator (SSE) has been used to determine the annual trend. fig 4a, fig. 4b., and fig. 4c show the annual trend of Minimum, Maximum and Average ambient temperatures using Sen's estimator. Results from Mann-Kendall test are discussed. The average seasonal trend for the 14 years is shown in fig. 3.

### 4.1.1 Seasonal Trend of Ambient temperature

Fig. 3 shows the mean seasonal trend for the minimum, maximum and average ambient temperature, and Table 2a shows the descriptive statistics for this. From the figure, it can be seen that the minimum ambient temperature variation for each year is fairly stable with a range of 1.25 and standard deviation of 0.46. The trend of the maximum ambient temperature has a wider range (5.46) and more deviation (1.78) compared to that of the minimum ambient temperature. The average ambient temperature trend clearly follows the same trend as the minimum and maximum ambient temperature, but with a higher range (3.03) and deviation (0.97) than the minimum ambient temperature and a lesser range and deviation than the maximum ambient temperature variation.

From our figure, February has the highest value of temperature for all trends (February and March for minimum ambient temperature trend). This corresponds to the months with little or no rainfall in the study area. The trend shows that the minimum ambient temperature has little or no effect on the temperature changes of the study area. This can be attributed to the almost uniform weather conditions in Calabar. As elaborated earlier, the deviation of the values of the minimum ambient temperature trend is small proving it's uniform and almost stable.

All seasonal temperature trends have their lowest values in the month of August. This corresponds to the month with the highest recorded rainfall in the study area. The maximum ambient temperature trend steadily rises from September till it attains its maximum value in February, and then drops steadily between March and August. This trend correlates slightly with the minimum ambient temperature variation. This shows that the variation of maximum ambient temperature has an effect on the variation of minimum ambient temperature.

The average ambient temperature plot follows the same trend as the minimum and maximum ambient temperature, but with a higher range and deviation than the minimum ambient temperature and a lesser range and deviation the maximum ambient temperature variation.

### 4.1.2 Annual Trend of Temperature

Results were gotten for the annual ambient temperature variations, fig 4a, fig 4b and fig 4c shows the trends for minimum, maximum and average ambient temperatures respectively. Sen's slope is represented in those figures also. From just mere observation, we can observe from fig 4b that the hottest year in the period is 2016, having a maximum ambient temperature of 304.83k (31.7°C). the year with the least maximum temperature can be observed to be 2006 with a value of 303.61K (30.46°C).

We can see that there is no direct observable relationship between the maximum and minimum temperature, as the years with the highest and least minimum temperature, observed in 2014 and 2010 were 296.99K; 23.8°C and 295.63K; 22.5°C respectively.

The M-K test has been adopted to analyze the significance of the trends of temperature along with Sen's slope ($Q$). results presented in table 5 shows that the minimum, maximum and average temperature trends have a Z-statistic value of -1.09, 2.53 and 0.33 respectively. We can discern from this with quintessence that the maximum and average temperature variation are both positively increasing because of their positive Z-value and the negative Z-value of the minimum temperature variation means it is reducing annually.

At 5% significance level, the maximum temperature trend rejected the null hypothesis $H_0$ and accepted the alternative hypothesis $H_1$ after its probability value (p-value) of this increasing series was found to be less than the significance level $\alpha$ [p-value (0.012) < 0.05 ($\alpha$)]. This shows that there is a trend in the series.

The p-values of the increasing (positive) average temperature and decreasing (negative) minimum temperature were 0.743 and 0.274 respectively; both greater than the 5% level of significance (0.05) [p-value (0.743)> 0.05 ($\alpha$) and p-value (0.274> 0.05 ($\alpha$) respectively]. This reveals that both trends (average and minimum ambient temperature) accepted the null hypothesis $H_0$ which says that there is no trend in the series.

Estimates from Sen's slope represented in fig 4a, fig 4b and fig 4c respectively all agreed with the Kendall Z and S statistic results. The Sen's slope ($Q$) magnitude for minimum, maximum and average temperature were calculated to be -0.018, 0.068 and 0.023 respectively; all agreeing with the Kendall Z-values of -1.09, 2.52 and 0.33 respectively. The Kendall Tau ($\tau$) represents these results by showing their relationships.

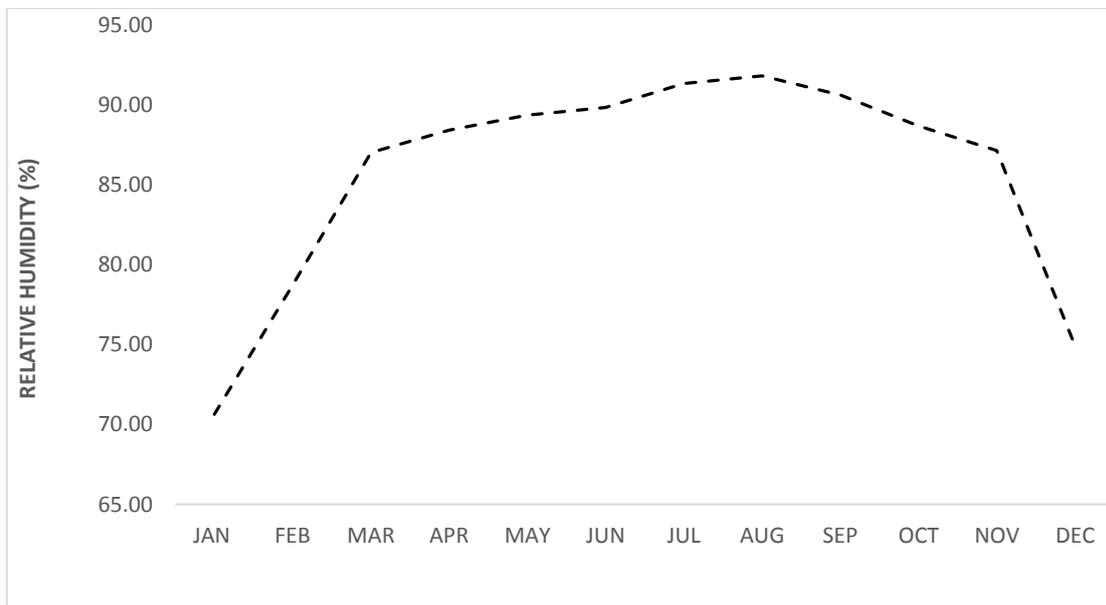

**Fig. 5** Mean seasonal relative humidity variation over the period of 14 years

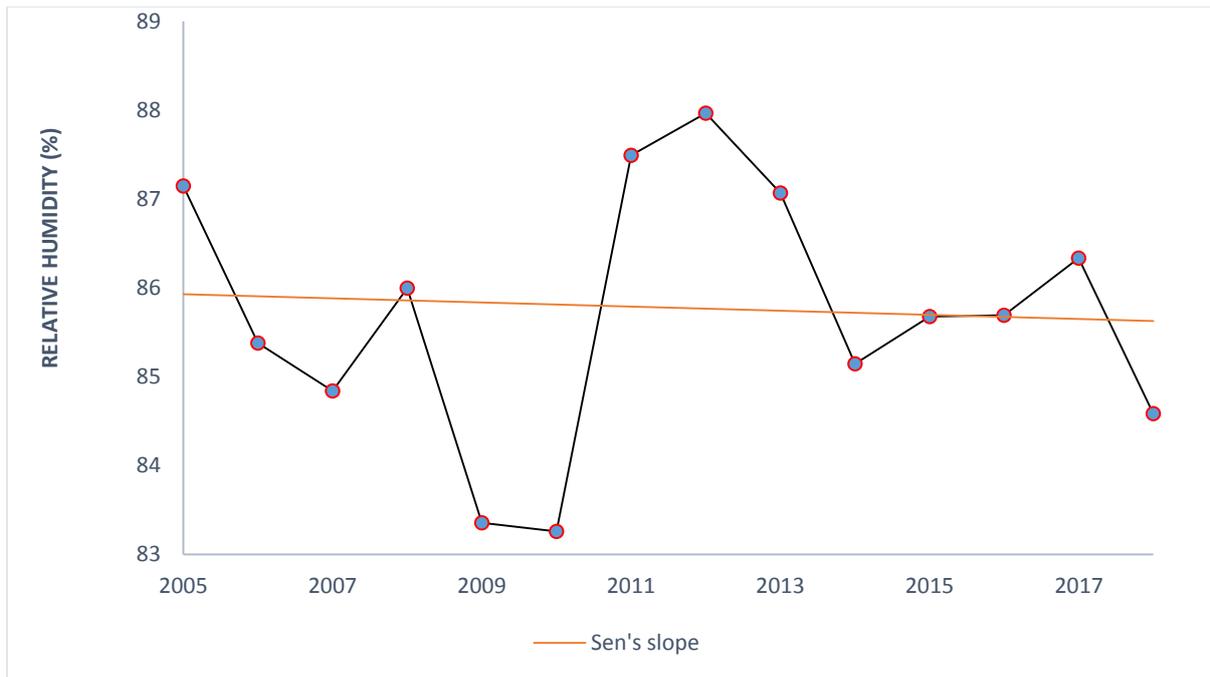

**Fig. 6** Sen's slope of relative humidity for the period of 2005-2018

## 4.2 RELATIVE HUMIDITY

### 4.2.1 Seasonal Trend of Relative Humidity

From fig. 5, it can be observed that the trend for Relative Humidity in a year follows a particular shape, having its peak between (June, July, August, September) and dropping during the latter parts of the year as well as the early parts. The relationship between the monthly ambient temperature (Fig. 3) and the monthly relative humidity trend (Fig. 5) shows a fairly strong negative coefficient of correlation -0.60 (Table 3). This implies that they have a strong negative relationship as the humidity is high when the ambient temperature is low for months in a year. The highest value of relative humidity corresponds to the lowest value of ambient temperature. This proves our negative correlation.

The relationship between ambient temperature and relative humidity can be demystified by the understanding of the inverse relationship between both parameters. An increase in ambient air temperature causes air to retain more water molecules, reducing their relative humidity and vice versa. This arises from the fact that for a lot of moisture is required to make hotter or warmer air saturated than colder air.

### 4.2.2 Annual Trend of Relative Humidity

Trend Analysis of Calabar in Cross River State has been done with 14 years' relative humidity data from 2005-2018. Mann-Kendall test and Sen's Slope Estimator has been used to determine the annual trend. Fig. 6 shows the annual trend of relative humidity using Sen's slope estimator. Results from Mann-Kendall test are discussed. The most humid year of the study was observed in 2012 (87.97%) while the year with the lowest relative humidity was observed in 2010 (83.26%).

The relative humidity fluctuations were analyzed and found to be decreasing annually after calculations revealed a Kendall Z-statistic value of -0.11. this was shown by the negative progressive trend. Close examination of results presented in table 5 shows negative S-statistic and Kendall Tau ($\tau$) Values.

Results have proven that the relative humidity variation is negative and decreasing but however, the probability value (p-value) of the series was calculated to be greater than the significance level ($\alpha$) [p-value (0.913)> 0.05 ($\alpha$)]. This means that for the annual relative humidity trend, one cannot reject the null hypothesis $H_0$, which says that there is "no trend" in the series.

Fig 6 graphically shows this decrease in trend via the Sen's slope (Q), this value is -0.023 which validates the results of the Kendall Z-statistic value (-0.11) of a reducing relationship. This is in contrast to the average ambient \temperature which had a positive value of Sen's slope for relative humidity (0.023). This proves their inverse relationship.

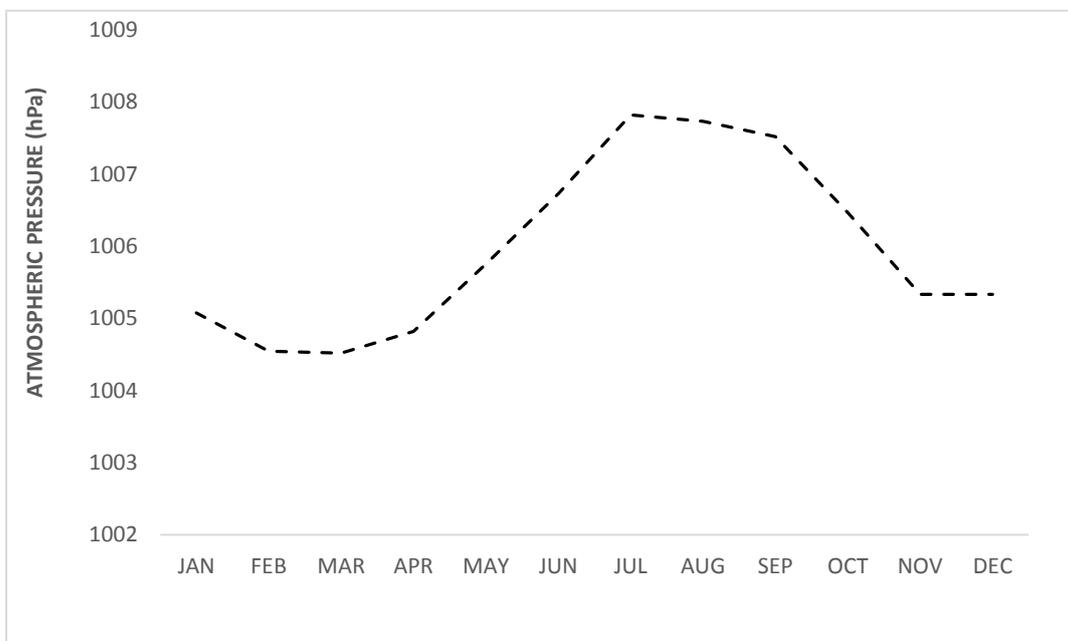

**Fig. 7** Mean seasonal atmospheric pressure variation over the period of 14 years

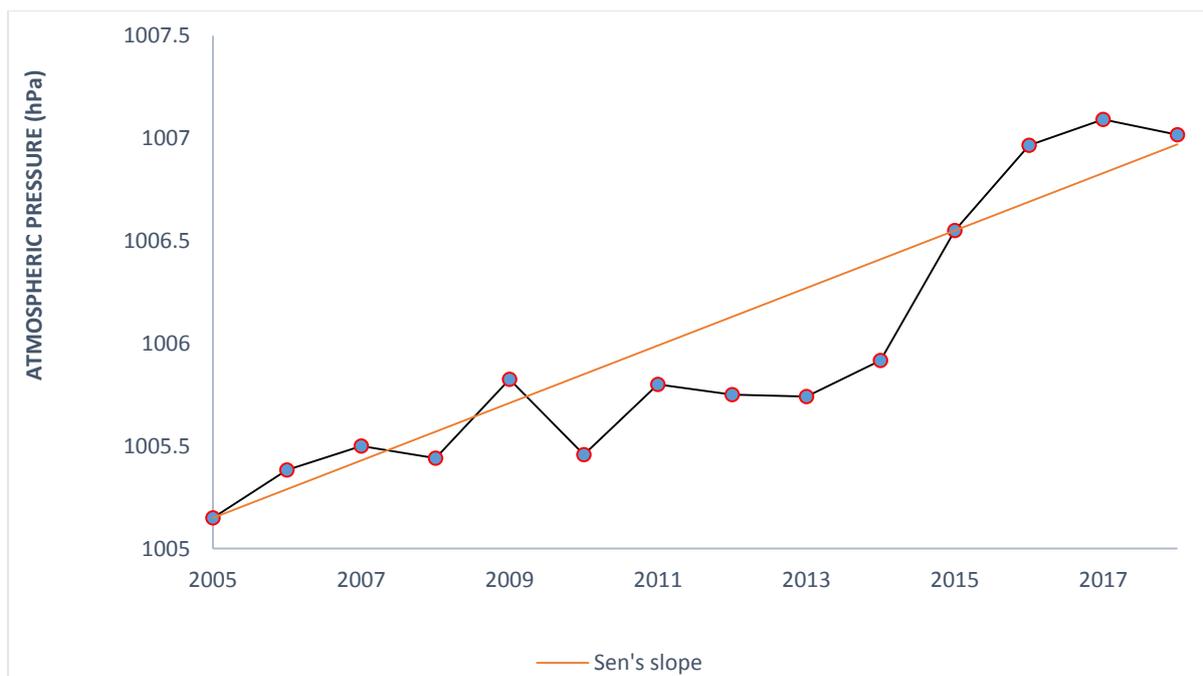

**Fig. 8** Sen's slope of atmospheric pressure for the period of 2005-2018

## 4.3 ATMOSPHERIC PRESSURE

### 4.3.1 Seasonal Trend of Atmospheric Pressure

Fig. 7 shows the average seasonal trend for atmospheric pressure. The trend peaks during the months of July, August and September. It is important to note that the coefficient of correlation between atmospheric pressure and relative humidity yielded a fairly strong positive correlation of 0.60, this can be explained by taking note of the highest values of atmospheric pressure, which were observed in the months of July (1007.82 hPa) and August (1007.73 hPa), which corresponds to the months with the highest measured relative humidity. This therefore means that the atmospheric pressure and relative humidity has a direct relationship as the increase in the water vapour content of the atmosphere increases the pressure on the troposphere.

The coefficient of correlation between atmospheric pressure and average ambient temperature yielded a very strong negative value of -0.97, this can be similarly interpreted by taking note of the lowest values of atmospheric pressure observed in the months of March (1004.53 hPa) and February (1004.55 hPa), which corresponded with the months with the highest average ambient temperature. This means that the atmospheric pressure has a strong indirect relationship with the average ambient temperature because the increase in atmospheric pressure causes a decrease in average ambient temperature and vice-versa.

### 4.3.2 Annual Trend of Atmospheric Pressure

Trend Analysis of Calabar in Cross River State has been done with 14 years' atmospheric pressure data from 2005-2018. Mann-Kendall and Sen's Slope Estimator has been used to determine the annual trend. Fig. 8 shows the annual trend of atmospheric pressure using Sen's estimator. Results from Mann-Kendall test are discussed. The year with the highest atmospheric pressure was observed to be 2017 (1007.09 hPa) and 2018 (1007.20 hPa). The lowest observed year was the earliest measured year (2005; 1005.15 hPa).

Atmospheric pressure variation has been deconstructed for the obtained years with the M-K test and the series shows a high Z-statistic value of 3.83. This means that atmospheric pressure has been increasing annually over all the years as the Kendall S-statistic and Tau values agree with this (table 5). These values were found to be the highest amongst all parameters analyzed. The increasing atmospheric pressure variation was checked to know the significance and the p-value of the series was found to be far less than the significance level ($\alpha$) [p-value (0.0001) $\ll$ 0.05 ($\alpha$)]. We can deduce with accuracy that there is 'a trend' in the atmospheric pressure series after the null hypothesis $H_0$ was rejected and the alternative hypothesis $H_1$ accepted.

The graphical representation of this significantly increasing atmospheric pressure trend has been shown in fig 8, showing the Sen's slope $Q$. Results from Sen's slope agrees with the positive results of the M-K test after having a slope ($Q$) of 0.140. These positive results are consistent for all results of atmospheric pressure as observed from table 5.

|  | Average Ambient Temperature | Relative Humidity | Atmospheric Pressure |
|---|---|---|---|
| **Average Ambient Temperature** | 1.00 | -0.59 | -0.97 |
| **Relative Humidity** | -0.59 | 1.00 | 0.60 |
| **Atmospheric Pressure** | -0.97 | 0.60 | 1.00 |

**Table 3.** Correlation of all meteorological parameters with each other, showing the significance of the relationship they have on each other's variation.

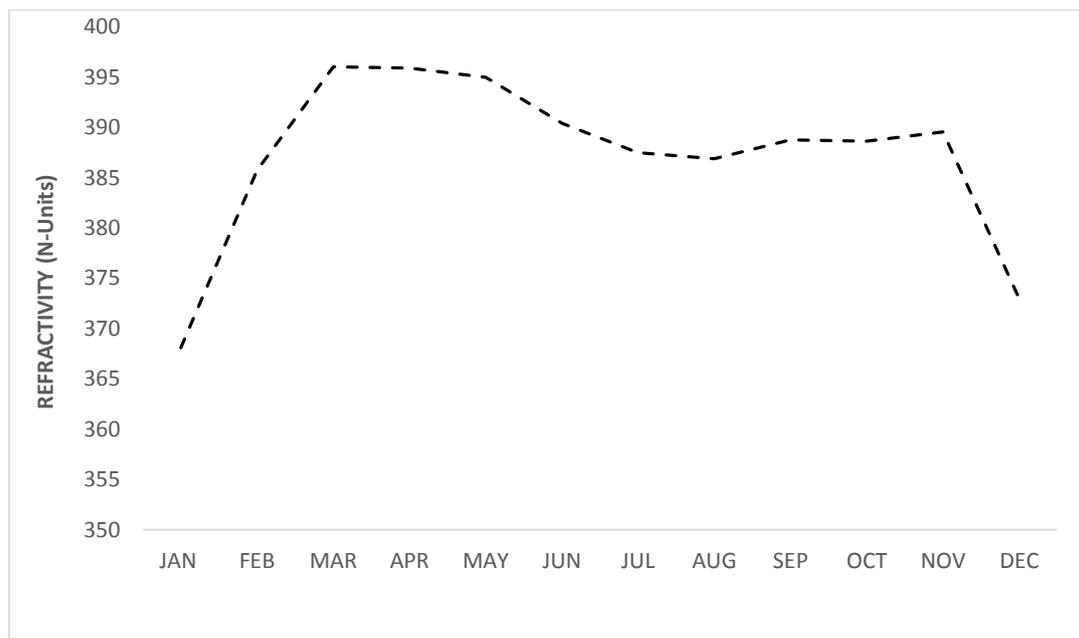

**Fig. 9** Mean seasonal refractivity variation over the period of 14 years

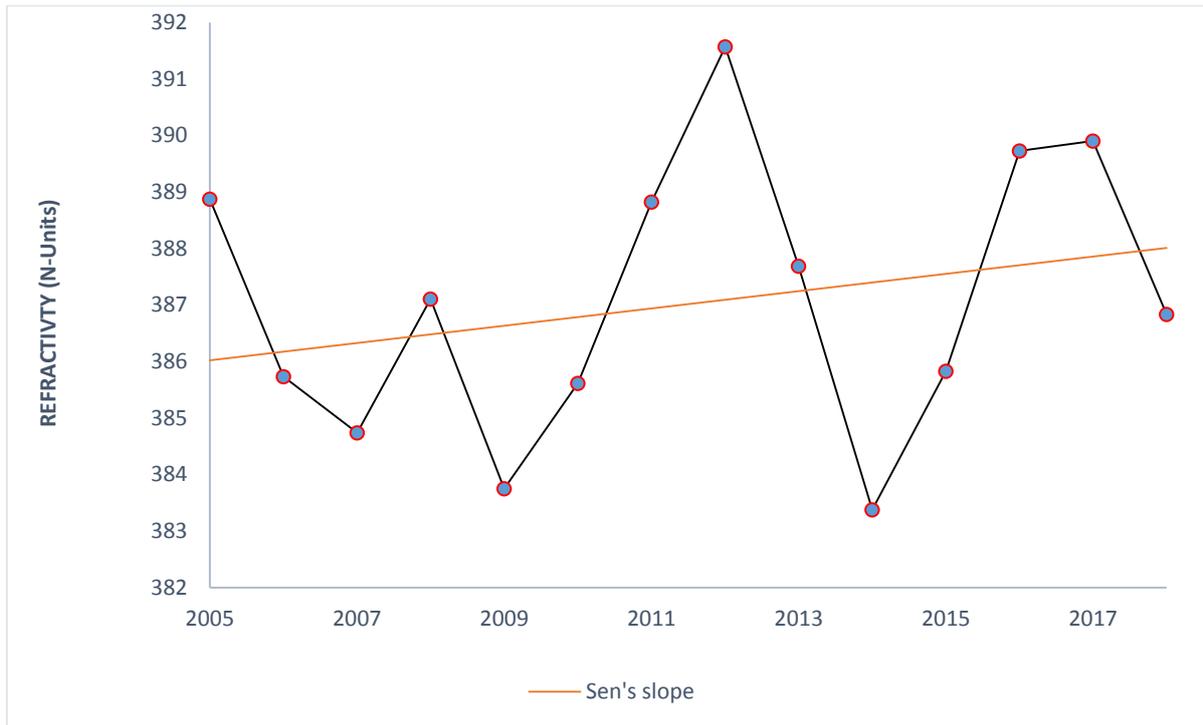

**Fig. 10** Sen's slope of refractivity for the period of 2005-2018

## 4.4 Refractivity

### 4.4.1 The Degree of Influence of Each Meteorological Parameter on Refractivity

A linear relationship between each meteorological parameter and radio refractivity has been obtained to determine the driving force behind the trend of radio refractivity. The slope and its error, the intercept and its error have all been represented in the linear regression equation describing the relationship. The LINEST function in Excel has been used to determine these values from the array of numbers by calculating the intercept normally, and returning additional regression statistics. The coefficient of correlation between each meteorological parameter has been found and represented in Table 3.

The linear relationship between refractivity and atmospheric pressure is very weak (equation 26), this is shown in Table 4 as the correlation coefficient is 0.25. Thus, the percentage contribution of atmospheric pressure to refractivity in this region is about 25%. The linear relationship shows a large percentage error in slope (111.47%) and intercept (189.81%) proving that atmospheric pressure cannot be used to accurately predict the variation of refractivity.

$$Refractivity\left(N_s\right) = \left(0.93 \pm 1.03\right)Pressure - 550.87 \pm 1045.62 \qquad (26)$$

The equation for the average ambient temperature (equation 27) also like the pressure does not have a strong coefficient of correlation with refractivity, from fig. 4, the value is 0.48. This shows a low effect on refractivity as average ambient temperature contributes to about 48% of refractivity variation in the region. The linear relationship shows a relatively smaller percentage error in slope (53.35%) and intercept (76.79%) compared to that of atmospheric pressure, but still relatively high compared to relative humidity.

$$Refractivity\left(N_s\right) = \left(4.22 \pm 2.25\right) Temperature - 881.37 \pm 676.76 \tag{27}$$

On the other hand, the correlation between radio refractivity and relative humidity is 0.76; showing that relative humidity is found to have contributed approximately 80% to variation of radio refractivity (equation 28). The percentage error in the slope (24.49%) and intercept (27.32%) is the smallest for all parameters, proving its accuracy. This implies that with the values of relative humidity the variation of refractivity in the region can be predicted with an accuracy of about 80%. This agrees with Agbo, et al., (2020).

$$Refractivity\left(N_s\right) = \left(1.30 \pm 0.32\right) Humidity - 275.54 \pm 27.32 \tag{28}$$

To have a better theoretical understanding of the meteorological parameter which contributes to the most variation to refractivity, we differentiate equation (4) partially with respect to refractivity elaborated from Agbo et al (2020). Differentiating equation (4) partially gives;

$$\partial N = 77.6 \frac{\left(T\partial P - P\partial T\right)}{T^2} + 3.75 \times 10^5 \frac{\left(T^2 \partial e - 2eT\partial T\right)}{T^4} \left(N - units\right)$$

simplified to;

$$\partial N = 77.6 \frac{\partial P}{T} - \left(77.6 \frac{P}{T^2} + 7.46 \times 10^5 \frac{e}{T^3}\right)\partial T + 3.73 \times 10^5 \frac{\partial e}{T^2}\left(N - Units\right) \tag{29}$$

Now we can theoretically determine which meteorological parameter has the most effect on the changes in refractivity by using equation (29) to solve for $\partial N$. Using the overall average values of $T = 300.28$K, $RH = 85.71\%$, $P = 1005.71$ hPa, $e = 30.71$ hPa, we obtain the change in refractivity to be;

$$0.258\partial P - 1.712\partial T + 4.137\partial e \tag{30}$$

From equation (30), it can be observed that the vapour pressure e, which is related to the relative humidity H, has the most effect on the gradient of radio refractivity. This theoretically proves our linear regression model for relative humidity in equation (28). Table 4 represents the represents the relationship between each meteorological parameter and radio refractivity.

| Parameters | Coefficient of correlation (R) | Slope | Error in slope | Percentage error in slope (%) | Y Intercept | Error in Y intercept | Percentage error in Y intercept (%) |
|---|---|---|---|---|---|---|---|
| Average Ambient Temperature | 0.48 | 4.22 | 2.25 | 53.35 | -881.37 | 676.76 | 76.78 |
| Relative Humidity | 0.76 | 1.30 | 0.32 | 24.49 | 275.54 | 27.32 | 9.92 |
| Atmospheric Pressure | 0.25 | 0.93 | 1.04 | 111.47 | -550.87 | 1045.62 | 189.81 |
| EPT | 0.95 | 1.59 | 0.16 | 9.7 | -180.93 | 55.60 | 30.73 |

**Table 4** Relationship between each meteorological parameter, equivalent potential temperature (EPT) and refractivity, showing the coefficient of correlation and the results from the linear regression.

### 4.4.2 Seasonal Trend of Refractivity

The average seasonal trend of refractivity for all years has been displayed in fig. 9. The trend slightly follows that of the seasonal relative humidity is fig. 5 as the highest values of refractivity in a year in measured in the wet months with high relative humidity, this proves our correlation coefficient between refractivity and relative humidity in table 4. and our linear relationship in equation (28). From Table 2a., the highest value of refractivity on average was in the month of March with 396.03 N-units and its lowest in January with 368.07 N-units. The trend is relatively stable throughout the wet months (March to November) which corresponds to the most humid months in fig. 5.

### 4.4.3 Annual Trend of Refractivity

Trend Analysis of Calabar in Cross River State has been done with calculated 14 years' refractivity data from 2005-2018. Mann-Kendall and Sen's Slope Estimator has been used to determine the annual trend. Fig. 10 shows the annual trend of refractivity using Sen's estimator. Results from Mann-Kendall test are discussed. The highest value of refractivity can be observed from table 2a. to be in 2012 (391.57 N-units) which corresponds with the year with the highest relative humidity, while the year with the lowest refractivity was observed in 20104 (383.38 N-units).

By the mere observation of fig 10, we can see that the annual variation of refractivity is increasing. This is proven by the analysis of the M-K test which shows a positive Kendall Z-statistic value of 0.76. this positive value as well as the Kendall Tau and S statistic values show a positive series, but this refractivity variation has been found to not be significant as the p-value is greater than the level of significance ($\alpha$) [p-value (0.443)> 0.05 ($\alpha$)]. This means that although refractivity is increasing over the years, it is not doing so with significance at 95% confidence level and hence the null hypothesis $H_0$ is accepted saying that there is "no trend" in the series. Table 5 shows the trend analysis results from the M-K test for all parameters.

Observed from fig 10, the positive Sen's slope is quickly discerned. The Sen's slope ($Q$) for the refractivity series is calculated to be positive (0.153). this is in agreement with the results from the M-K test having a positive Z-value of 0.76, showing that although refractivity of waves is increasing, it is not doing so with much significance.

| Variables | Kendall's Tau | Mann Kendall's Statistic (S) | VAR(S) | Test Statistic (Z) | p-value (Two-tailed) | Alpha (α) | Sen's slope (Q) | Test Interpretation |
|---|---|---|---|---|---|---|---|---|
| Minimum Ambient Temperature | -0.231 | -21.00 | 333.667 | -1.09 | 0.274 | 0.05 | -0.018 | $H_0$ ↓ |
| Maximum Ambient Temperature | 0.516 | 47.00 | 333.667 | 2.52 | 0.012 | 0.05 | 0.068 | $H_1$ ↑ |
| Average Ambient Temperature | 0.077 | 7.00 | 333.667 | 0.33 | 0.743 | 0.05 | 0.023 | $H_0$ ↑ |
| Relative Humidity | -0.033 | -3.00 | 333.667 | -0.11 | 0.913 | 0.05 | -0.023 | $H_0$ ↓ |
| Atmospheric Pressure | 0.780 | 71.00 | 333.667 | 3.83 | 0.0001 | 0.05 | 0.140 | $H_1$ ↑ |
| Refractivity | 0.165 | 15.00 | 333.667 | 0.77 | 0.443 | 0.05 | 0.153 | $H_0$ ↑ |
| EPT | 0.099 | 9.00 | 333.667 | 0.44 | 0.661 | 0.05 | 0.055 | $H_0$ ↑ |
| Field Strength | 0.648 | 59.00 | 333.667 | 3.18 | 0.001 | 0.05 | 0.114 | $H_1$ ↑ |

**Table 5** Results from Mann Kendall's trend test and Sen's slope for each meteorological parameter, calculated equivalent potential temperature (EPT), refractivity and field strength showing the nature of the trend [increasing (↑) or decreasing (↓)], and results from the hypotheses, represented by $H_0$ (No trend in the series) and $H_1$ (There is a trend in the series)

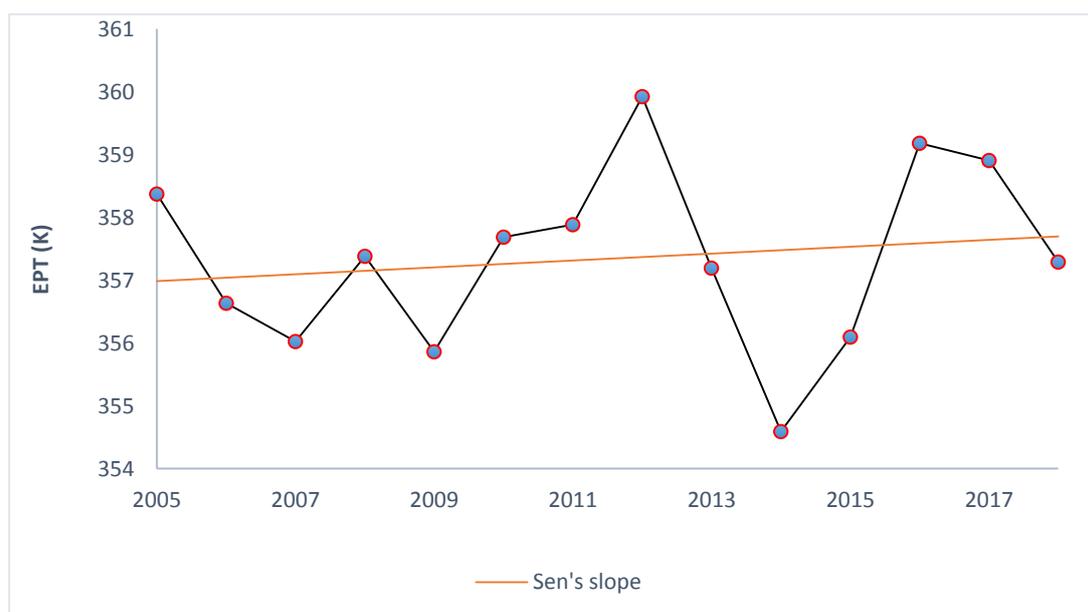

**Fig. 11** Sen's slope for Equivalent potential temperature for the period of 2005-2018

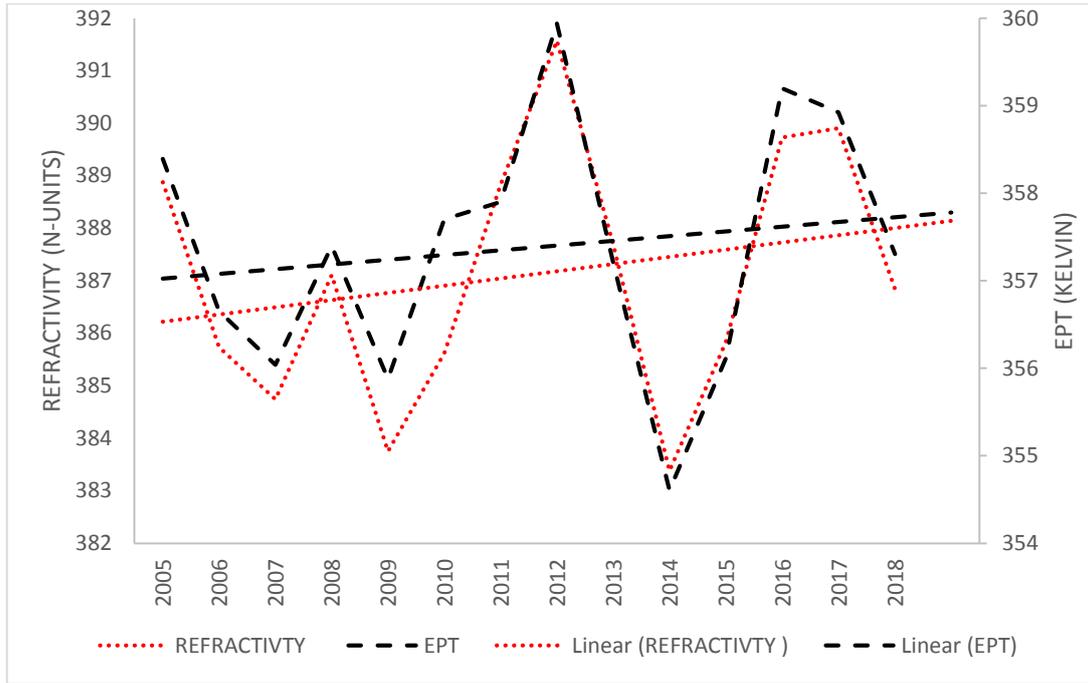

**Fig 12.** Comparison between annual trend of refractivity and EPT

### 4.5 Equivalent Potential Temperature

Adopting the M-K test and SSE to analyze the annual variation of EPT, fig. 11 shows the trend using SSE. The variation follows that of refractivity in (10); the Kendall Z-statistic value of 0.44 shows proof that a slightly positive trend has been discovered, but just as in the case of refractivity, the increase was not significant as the p-value was higher than the significance (α) [p-value (0.661)> 0.05 (α)]. This agrees with the null hypothesis ($H_0$), that there is no significant trend in the series.

The Sen's slope shows a slightly positive value of 0.055 agreeing with the Kendall's positive Z-value of 0.41. These results therefore mean that the variation (slight increase) of the EPT has a direct effect or is directly related to the variation of refractivity.

The link between refractivity and the EPT has been discerned. Just as in the case of all meteorological parameters, a linear relationship has been found between the two variables after it was discovered that there exists a very strong positive correlation between them. The coefficient of correlation between refractivity and EPT is found to be 0.95; this shows a very strong relationship between both trends proven from fig 12.

The strong correlation coefficient shows that the EPT for a pseudo-adiabatic process can be used to predict the refractivity variation with an accuracy of about 95%.

$$Refractivity\left(N_s\right) = \left(1.59 \pm 0.16\right) EPT - 180.93 \pm 55.60 \qquad (31)$$

The percentage error in the slope from the linear regression representation in equation () is (9.7%) and intercept (30.73%), this shows the accuracy of the relationship.

The trends for refractivity (fig. 10) and EPT (fig. 11) are both shown in fig. (12). The purpose of this is to depict the direct relationship between both variables. The linear relationship can be clearly seen as the trend is almost similar.

This means that an unsaturated air parcel would rack-up a temperature directly proportional to the variation of refractivity when lifted adiabatically to condense and then brought down adiabatically to 1000hPa. Again, this proves the relationship between refractivity and water vapour (moisture), which is in agreement with Gao et al. (2008), whose results showed that that the temperature has little or no effect on refractivity variation, unlike water vapour (relative humidity). Howbeit, in their study, the EPT wasn't considered, and from our results, its variation shows a large similarity with refractivity and hence, a large contribution from water vapour.

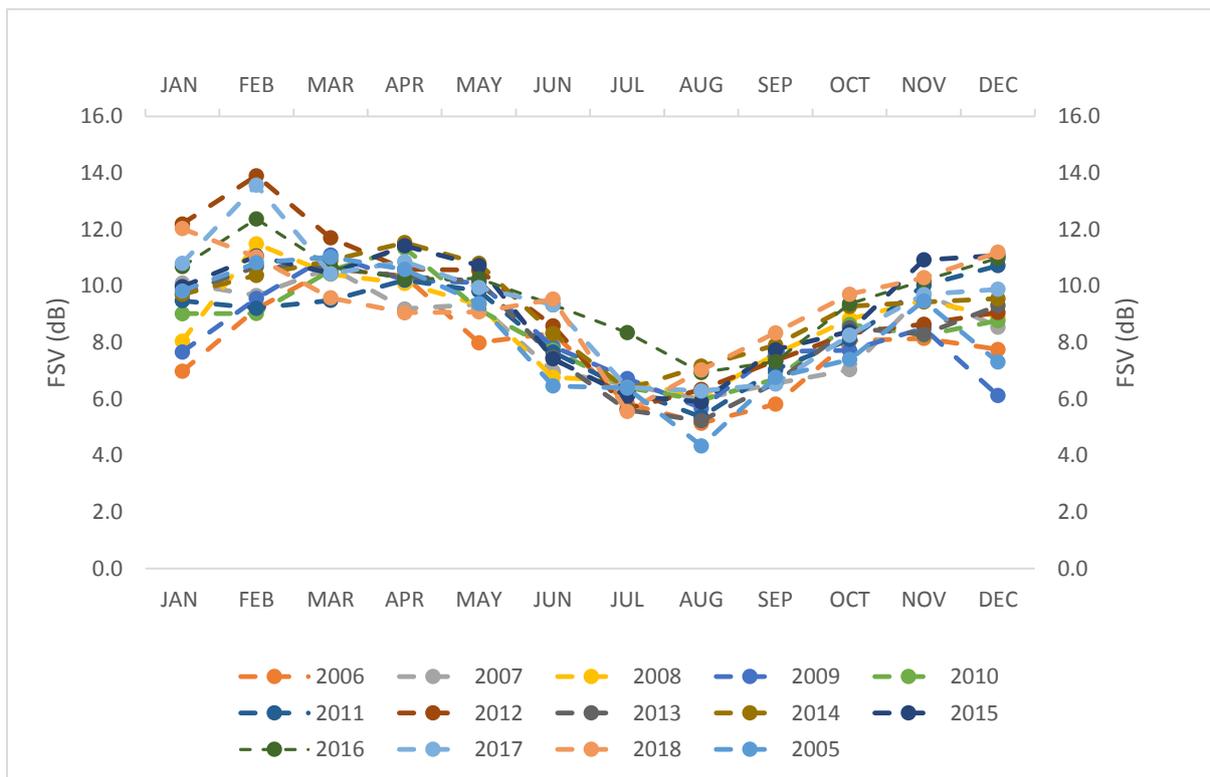

**Fig. 13** Seasonal field strength variation over the period of 14 years

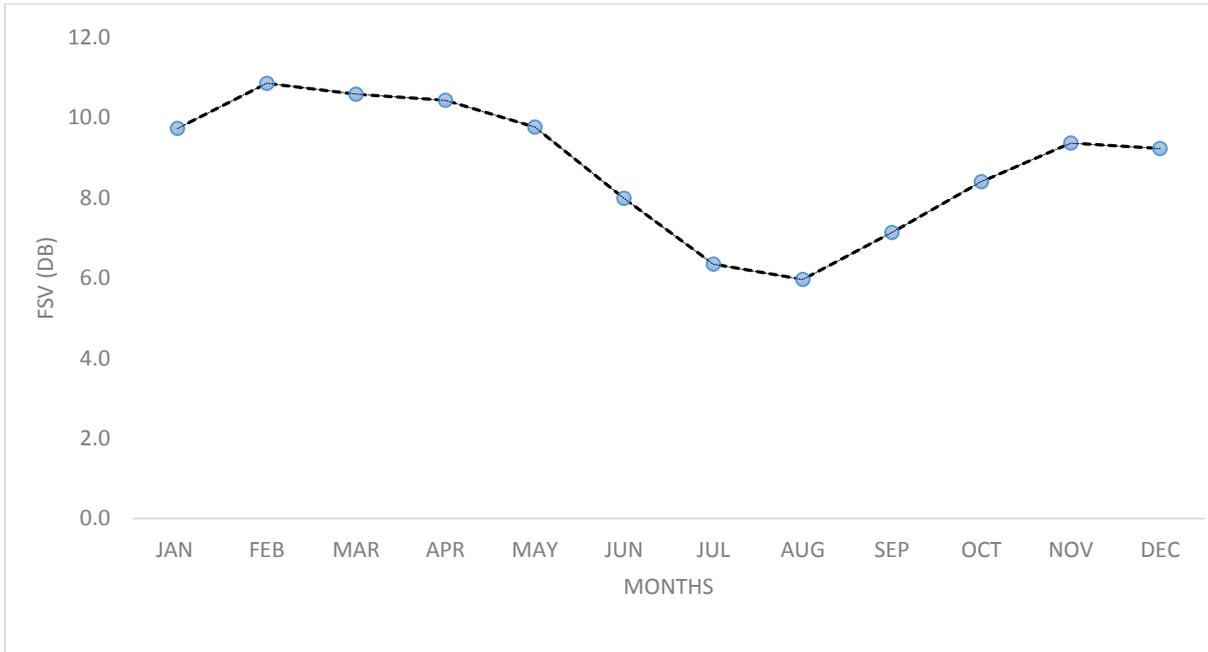

**Fig. 14** Mean seasonal field strength variation over the period of 14 years

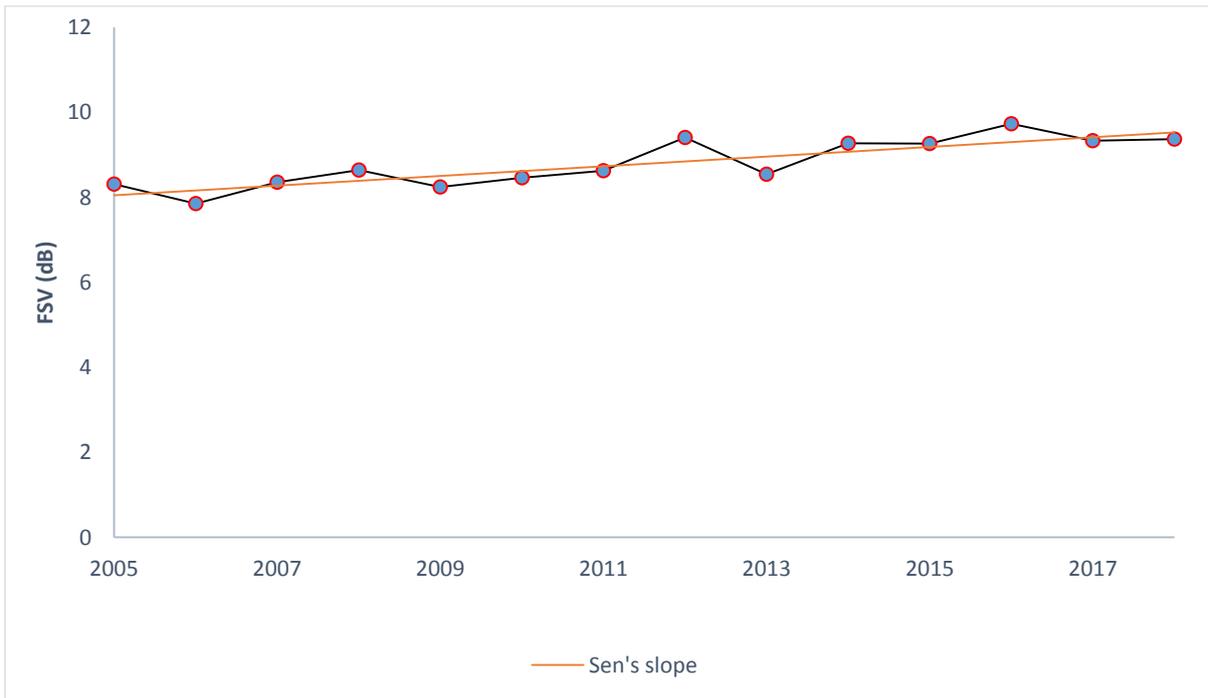

**Fig. 15** Sen's slope of field strength for the period of 2005-2018

**4.6 Field Strength Variability (FSV)**

**4.6.1 Seasonal Trend of Field Strength**

Fig. 13 and 14 shows the seasonal trends of field strength. The Field Strength variability was observed to have higher values in the months with lower temperature and lower values in the months with higher relative humidity, unlike the refractivity which correlates directly with humidity; the field strength is always low during the humid months. The highest value of field strength throughout the months was observed in the month of February to be 10.9 dB; this corresponds to the month with the highest temperature. The lowest value of field strength can be observed to be in the month of August (6.0 dB); this corresponds to the month with the highest relative humidity. This means that the existence of water vapour in the atmosphere increases the refractivity, decreases the strength of the field and vice-versa. This shows that the field strength is dependent on the moisture content in the atmosphere, which agrees with Tanko, et al., (2019) results for northern Nigeria.

**4.6.2 Annual Trend of Field Strength**

Trend Analysis of Calabar in Cross River State has been done with 14 years' calculated Field Strength data from 2005-2018. Mann-Kendall and Sen's Slope Estimator has been used to determine the annual trend. Fig. 15 shows the annual trend of atmospheric pressure using Sen's estimator. Results from Mann-Kendall test are discussed. With an average of 9.73dB, 2016 was observed to have the highest Field Strength, the lowest field strength value was observed 10 years prior in 2006 (7.86dB).

The outcome of the non-parametric M-K test after being used to analyze the variability of field strength reveals a Z-value of 3.18. meaning that FSV is increasing annually and this variation has been deduced to be significantly increasing after its computed p-value was observed to be far less than the significance level ($\alpha$) of 5% [p-value (0.0001) <0.05($\alpha$)]. This agrees with the alternative hypothesis $H_1$ which says that there is a trend in the series contrary to the alternative hypothesis $H_0$. This shows that the field strength is significantly increasing as the years go by as can be observed from the calculated values.

The Sen's slope ($Q$) of the field strength can be observed from fig 15 and is in agreement with the results from the M-K test showing a positive and significantly increasing trend, The Sen's slope $Q$ is positive and calculated to be 0.114.

**5 Conclusion**

We can conclude from both Mann- Kendall and Sen's Slope that there is tendency of increment with significance in the annual trend of atmospheric pressure, maximum ambient temperature, field strength. However, the annual trend of refractivity, EPT, and average ambient temperature are also increasing, but without significance. Similarly, the annual trend of minimum ambient temperature and relative humidity are decreasing without significance. This clearly shows that the atmospheric pressure and the maximum ambient temperature in the region is increasing and should be monitored.

Our study shows that the radio refractivity is directly affected by the water vapour content of the atmosphere as the months with the highest relative humidity are observed to have the highest refractivity; The coefficient of correlation, linear regression and partial differentiation of refractivity has been used to prove this experimentally and theoretically respectively. On the contrary, the field strength has an inverse relationship with relative humidity as the months with the highest field strength were recorded in the dry months., with its lowest coming in the month of August which is the most humid month, which can be explained from the correlation of all meteorological parameters with themselves (fig. 3) The novel result on the relationship between refractivity and EPT has been discerned, and results show that there is a strong positive relationship between the two variables (correlation coefficient R = 0.95) . This similarly proves the close relationship between refractivity and water vapour since the EPT is directly related to the temperature at the lifting condensation level.

## Acknowledgments

We thank the referees for their positive and enlightening comments and suggestions, which have greatly helped us in making improvements to this article. In addition, the authors acknowledges the Nigerian Meteorological Agency (NiMet), Calabar for providing the necessary data used in this study.